\documentclass[12pt,preprint]{aastex}
\newif\ifpp\pptrue
\newcommand{\bc}{\begin{center}}
\newcommand{\ec}{\end{center}}
\newcommand{\be}{\begin{equation}}
\newcommand{\ee}{\end{equation}}
\newcommand{\bi}{\begin{itemize}}
\newcommand{\ei}{\end{itemize}}
\newcommand{\bn}{\begin{enumerate}}
\newcommand{\en}{\end{enumerate}}
\newcommand{\etal}{{\it et al.}}
\ifpp
\else
\fi
\singlespace
\begin{document}

\title{Distant Supernovae and the Accelerating Universe}
\author{Edward L. Wright, UCLA Astronomy}

\noindent
{
ABSTRACT:
The observation of SN 1997ff at redshift 1.7 has been 
claimed$^{\ref{ref:1997ff}}$ to refute alternative models such as grey dust or 
evolution for the faintness of distant supernovae, leaving 
only an accelerating Universe as a viable model. However, a 
very simple one parameter evolution model, with the peak 
luminosity varying as an exponential function of cosmic 
time, converts the flux vs. distance law of the critical 
density matter-dominated model into that of the concordance 
$\Omega = 0.3$ flat vacuum-dominated model with an error no 
larger than 0.03 mag over the range 0-2 in redshift. A grey 
dust model that matches this accuracy can easily be 
contrived but it still fails by overproducing the far-IR 
background or distorting the CMB. Models that involve 
oscillation between photons and axions could emulate an 
exponential function of cosmic time without violating these 
background constraints. Clearly a better and well-tested 
understanding of the Type Ia supernova explosion mechanism 
and the origin of the correlation between the decay rate and 
luminosity is needed before any effort to reduce statistical 
errors in the supernova Hubble diagram to very small levels.
}

\ifpp\else \newpage \fi

\begin{figure}[tp]
\plotone{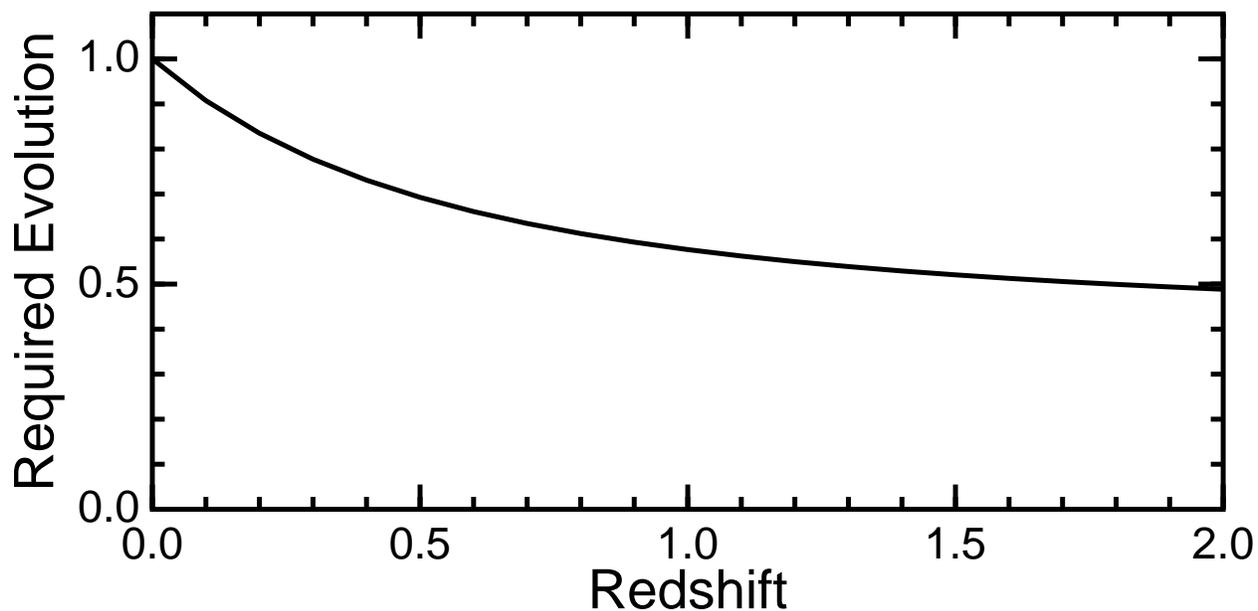}
\caption{The evolution required to make an $\Omega_M = 1$
model look like an $\Omega_M = 0.3,  \Omega_\Lambda = 0.7$
model in the supernova Hubble diagram.
\label{fig:Lvsz}}
\end{figure}

The observations of Type Ia SNe are only determining the luminosity
distance {\it vs.} redshift law, $D_L(z;P)$, where $P$ is a set of
cosmological parameters, typically $\Omega_M$ and 
$\Omega_\Lambda$.  Obviously if the intercept of the
supernova luminosity vs decay rate relation evolves$^{\ref{ref:ELD}}$
following the relation
\be
L(z) = (D_L(z;P^\prime)/D_L(z;P))^2
\label{eq:Lz}
\ee
then the true cosmological parameters would be $P^\prime$ even
though $D_L(z;P)$ fits the data well under the assumption
of no evolution.  If I evaluate Eqn(\ref{eq:Lz}) with 
$P^\prime = \{\Omega_M = 1,  \Omega_\Lambda = 0\}$
being the Einstein - de Sitter critical density matter dominated
model and $P = \{\Omega_M = 0.3,  \Omega_\Lambda = 0.7\}$
being the concordance model, then Figure \ref{fig:Lvsz} is obtained.

\begin{figure}[bp]
\plotone{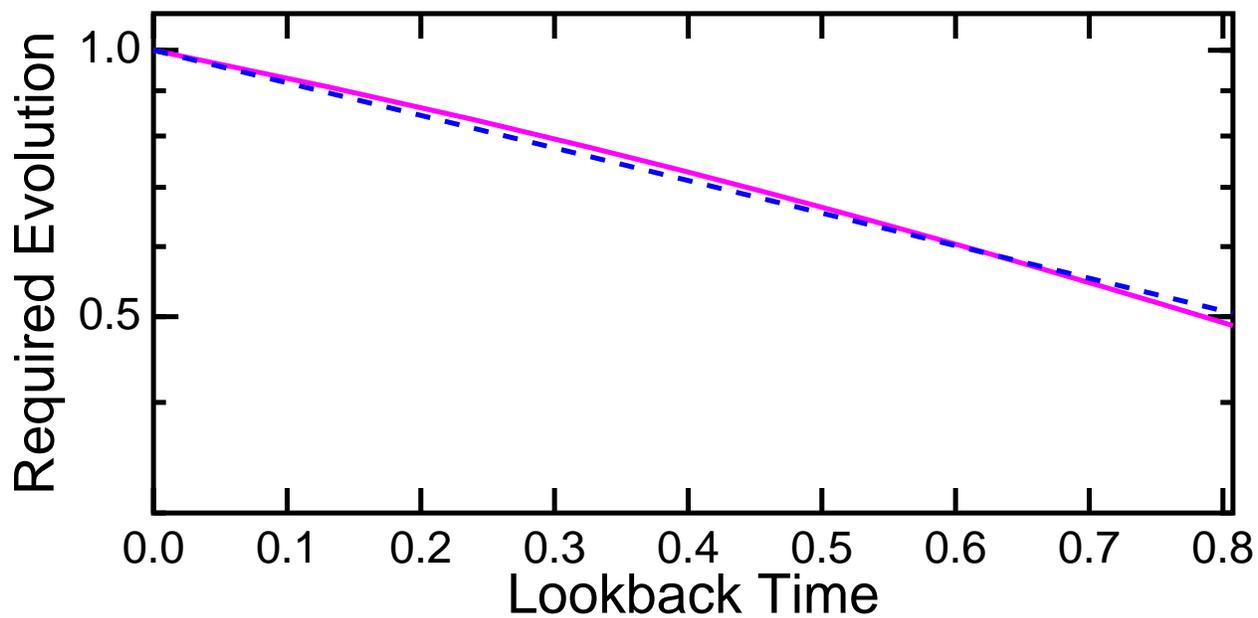}
\caption{The evolution required to make an $\Omega_M = 1$
model look like an $\Omega_M = 0.3,  \Omega_\Lambda = 0.7$
model in the supernova Hubble diagram.  The dashed 
line is an exponential function of cosmic time.
\label{fig:Lvst}}
\end{figure}

\begin{figure}[tbp]
\plotone{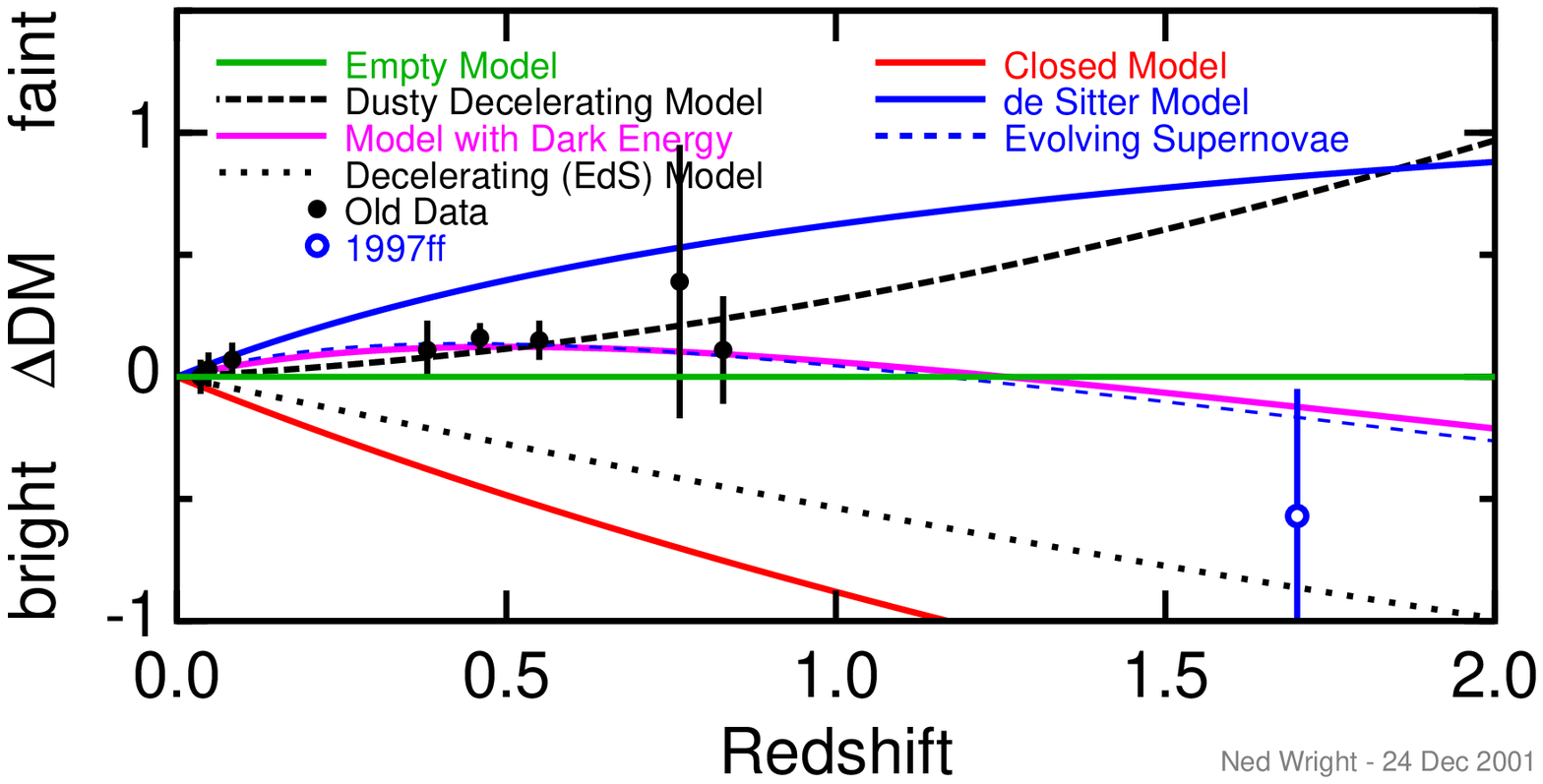}
\caption{The Type Ia SNe data plotted relative to an empty
Universe Milne model.  The exponential evolution matches
the concordance model to $\leq 0.028$ mag for all 
$z \in [0,2]$.   The ``dusty decelerating model'' has a constant comoving
dust density.  A dust model with a constant physical dust density
matches the exponential evolution.
\label{fig:DLvsz}}
\end{figure}

This required luminosity evolution is monotonic,
which contradicts the claim that evolution would have to reverse
itself to match the accelerating now but decelerating in the past
pattern of the concordance model.  Furthermore, if I plot the 
required evolution on a log-linear plot as a function of cosmic
time, I get the very nearly straight solid line on
Figure \ref{fig:Lvst}.  Obviously an exponential function of
cosmic time [the dashed line] provides a nearly exact
match to the required evolution.

Figure \ref{fig:DLvsz} shows various models including the
exponential function of cosmic time evolution model compared
to supernova data$^{\ref{ref:1997ff},\ref{ref:SCP99}}$.
Clearly both the concordance model
and the exponential evolution model fit the data very well --
actually too well -- with $\chi^2$ per degree of freedom much less
than 1.  The best fit $\Omega_\Lambda$ is $0.73$ with
$\chi^2 = 1.46$ for 8 df.
The evolution model is a slightly better fit, with 
$\chi^2 = 1.24$ for 8 df at $\alpha = 1.28$.  This evolution
model is equivalent to the axion model
in Eq(\ref{eq:axion}) below with $\beta=0$ \& $F=0$.


Given that the required evolution is monotonic, there is no
difficulty in producing this effect using grey dust.  One need
only make the physical dust density constant to give the
appearance of an exponential function of cosmic time evolution model.
Then the comoving dust density must vary like $(1+z)^{-3}$
which is certainly possible although it has no obvious connection to
the evolution of the star formation rate [the Madau curve].

Models where light -- but not millimeter waves -- can oscillate into
axions$^{\ref{ref:axion}}$ and thus lead to fainter distant supernovae, can 
also fit the data well.  But the standard 1 axion asymptote with 2/3 of the 
light surviving is not enough dimming of the distant supernovae if the
true model is an EdS Universe with $\Omega_M = 1$.   Letting the 
observed flux vary like
\be
\mbox{Flux} =
\frac{L}{4\pi D_L^2} [(1-F) \exp(-\alpha H_\circ \int_0^z (1+z)^\beta dt) + F]
\label{eq:axion}
\ee
one gets a best fit in an EdS Universe of $\chi^2 = 3.99$ with $\alpha = 7$
if $F$ is fixed$^4$ at
2/3 and $\beta$ is fixed at 1.  But with $\alpha$, $\beta$ and $F$ all free,
the best fit occurs when $\alpha = 2.61$, $F = 0.60$ and $\beta = 3.80$, giving
$\chi^2 = 0.85$ with 6 df.  While this modified axion model is the best fit,
and the optimum $F$ is quite close to the expected 2/3,
the model introduces many new parameters and invokes at least one tooth fairy
when it claims that $< 0.01\%$ of the CMB photons convert to axions while
converting 40\% of the optical photons.

\newpage

\bc
CONCLUSIONS
\ec
\vspace*{-15pt}
\bi
\item The Hubble diagram of distant Type Ia SNe is well fit by either
a vacuum-dominated accelerating Universe, or by exponential function 
of cosmic time evolution model in an $\Omega_M = 1$ Universe.
\item Both fits are better than would be expected for the quoted errors.
\item Both models require that {\it a priori} unlikely effects be added to 
the standard model.  The tooth fairy counts are equal.
\item One should wait for confirmation of the accelerating Universe
by independent means such as CMB data before committing a large
amount of resources to measuring the fine details of the SNe Hubble 
diagram.
\ei

\ifpp\else \newpage \fi

\bc
REFERENCES
\ec
\vspace*{-15pt}
\bn
\item Riess, A. \etal. 2001, \apj, 560, 49-71. 
``The Farthest Known Supernova: Support for an Accelerating Universe and a
Glimpse of the Epoch of Deceleration''
quote from abstract: ``It is inconsistent with grey dust or simple
luminosity evolution\ldots''
\label{ref:1997ff}
\item Regos, E., Tout, C., Wickramasinghe, D., Hurley, J. \& Pols, O. 2001,
astro-ph/0112355, ``Could Edge-Lit Type Ia Supernovae be Standard Candles''
quote from abstract: ``we find a systematic shift in this relation that would
make distant SNe Ia fainter than those nearby''
\label{ref:ELD}
\item Perlmutter, S. \etal\ 1999, \apj, 517, 565-586. 
``Measurements of Omega and Lambda from 42 High-Redshift Supernovae''
\label{ref:SCP99}
\item Csaki, C., Kaloper, N. \& Terning J. 2001 hep-ph/0111311
``Dimming Supernovae without Cosmic Acceleration''
\label{ref:axion}
\en

\end{document}